\documentclass[twocolumn,twoside,slac_two]{revtex4}
\usepackage{graphicx}
\usepackage{fancyhdr}
\newcommand{\gray}{\mbox{$\gamma$-ray}}
\newcommand{\grays}{\mbox{$\gamma$-rays}}

\newcommand{\Fermic}{\emph{Fermi}}
\newcommand{\Fermi}{\Fermic\ }
\newcommand{\FermiLATc}{\Fermic-LAT}
\newcommand{\FermiLAT}{\FermiLATc\ }

\newcommand{\lapp}{\ensuremath{\stackrel{\scriptstyle <}{{}_{\sim}}}}
\newcommand{\gapp}{\ensuremath{\stackrel{\scriptstyle >}{{}_{\sim}}}}

\newcommand{\degr}{$^{\circ}$}

\newcommand \apj         {ApJ}
\newcommand \apjs        {ApJS}

\begin{document}

\title{The First \FermiLAT Catalog of Sources Above 10 GeV\footnote{
The material reported in this proceedings is based on the full catalog paper which is currently in
preparation within the \FermiLAT\ collaboration.}
}

%

\author{D. Paneque}
\affiliation{Max Planck Institute for Physics, Munich, 80805, Germany}
\author{J. Ballet}
\affiliation{Laboratoire AIM, CEA-IRFU/CNRS/Universit\'e Paris Diderot, Service d'Astrophysique, CEA Saclay, 91191 Gifsur Yvette, France}
\author{T. Burnett}
\affiliation{Department of Physics, University of Washington, Seattle, WA 98195-1560, USA}

\author{S. Digel}
\affiliation{SLAC National Accelerator Laboratory, Menlo Park, CA, 94025, USA}

\author{P. Fortin}
\affiliation{Fred Lawrence Whipple Observatory, Harvard-Smithsonian Center for Astrophysics, Amado, AZ 85645, USA}
\author{J. Knoedlseder}
\affiliation{CNRS, IRAP, F-31028 Toulouse Cedex 4, France}

\author{\vspace{0.2cm} on behalf of the \FermiLAT collaboration}


\begin{abstract}
We present a catalog of $\gamma$-ray sources at energies above 10 GeV based on
data from the Large Area Telescope (LAT) accumulated during the first
3 years of the {\it Fermi Gamma-ray Space Telescope} mission.  This catalog complements the Second \FermiLAT Catalog, which
was based on 2 years of data extending down to 100 MeV and so included
many sources with softer spectra below 10~GeV. The First \Fermic-LAT Catalog of $>$10~GeV sources (1FHL) has
514 sources,
and includes their locations, spectra,
a measure of their variability, and associations
with cataloged sources at other wavelengths. We found that 449 (87\%)
could be associated with known sources,  of which 393 (76\% of the 1FHL sources) are
active galactic nuclei.  We also highlight the subset of the 1FHL
sources that are the best candidates 
for detection at energies above 50 GeV with ground-based $\gamma$-ray observatories. 

\end{abstract}

\maketitle

\thispagestyle{fancy}


\section{Introduction}

Many more \gray\ sources are known in the GeV range than above 100 GeV.
The ``\Fermi Large Area
Telescope Second Source Catalog'' \cite[hereafter 2FGL,][]{LAT_2FGL},
reports 1873 sources detected at energies above 100~MeV in the first 2 years of
\Fermi\ science operations. On the other hand, as reported in the TeVCat
catalog\footnote{\url{http://tevcat.uchicago.edu/}} version 3.400,
only 105 sources have been detected with ground-based \gray\
instruments above 100 GeV (VHE)\footnote{Including recently
  announced (but not yet published) VHE detections the number is 143.}, which is substantially
fewer than the number of objects reported in 2FGL. 

In this manuscript we briefly summarize the ongoing effort within the
\FermiLAT\ collaboration to produce the ``The First \FermiLAT
Catalog of Sources Above 10 GeV'', designated 1FHL. 
The lower limit of 10 GeV is a good
compromise between photon statistics and proximity to the energy range
of the ground-based \gray\ instruments. 
We report on the location, spectral and variability properties of the
514 sources significantly detected above 10 GeV. Many of these sources are already included in the 2FGL catalog, although in that catalog their characterization is dominated by the much larger numbers of \grays\ in the energy range 0.1--10 GeV. Consequently, the characteristics of the sources at the highest \FermiLAT\ energies might be overlooked because of the lower statistical weight of the most energetic (but less abundant) \grays. 
With its focus on the high-energy data, the 1FHL catalog complements
the 2FGL catalog, and is meant to be a sort of ``bridge catalog''
between the 2FGL and TeVCat.

\section{\FermiLAT Sources Above 10 GeV}

The \FermiLAT is a $\gamma$-ray telescope operating from $20$\,MeV to
$>300$\,GeV. The instrument is a $4 \times 4$ array of identical
towers, each one consisting of a tracker (where the photons are
pair converted) and a segmented calorimeter (where the energies of the
positron-electron pairs are measured). The tracker is covered
by an anti-coincidence detector to reject the charged-particle
background. Further details on \FermiLAT and its performance can be
found in \cite{LAT09_instrument} and \cite{LAT12_calib}.

In this work we utilize \emph{Pass7 Clean} events with energies in the range 10--500 GeV,
detected from
the beginning of science operations (2008 August 4, MET 239557447)
to 2011 August 1 (MET 333849586), covering very nearly 3 years\footnote{Mission Elapsed Time (MET) starts at 00:00 UTC on 2001 January 1 and does not include leap seconds.}.

As for the 2FGL analysis, source detection and characterization began with the assembly of a list of `seeds', candidate sources that were selected for input to the likelihood analysis chain. The seeds were used as input to standard maximum likelihood analysis to jointly optimize the spectral parameters of the candidate sources and to judge their overall significances (\S~\ref{SpectralAnalysis}).  This analysis was implemented in a similar way to the 2FGL analysis.  In the final step of the analysis we also searched for candidate counterparts of these 1FHL sources with sources in previous LAT catalogs and sources in known $\gamma$-ray emitting classes at other wavelengths (\S~\ref{SrcAssociations}) using the same methods as for the 2FGL source associations.
We put special focus on the characterization of the sources associated with
active galactic nuclei (AGNs), which constitute the majority of the
catalog (\S~\ref{FHL_AGNs}), as well as on the identification of candidate
sources for VHE detection with current ground-based \gray\
instrumentation (\S~\ref{GoodVHE}).

\subsection{Spectral Analysis}
\label{SpectralAnalysis}

The procedure we followed was similar to what was done for 2FGL.
Starting from the list of seeds 
we divided the sky into a number of Regions of Interest (RoI) covering
all source seeds; 561 RoIs were used for 1FHL.
Each RoI extends 2\degr beyond the sources it contains in order to
cover the entire PSF as well as allow the background diffuse emission to be 
well fit.
Because the spatial resolution is good above 10 GeV, there is little
cross-talk between sources or between RoIs, so global convergence
was relatively easy to achieve. 

Over the relatively narrow range 10 to 500 GeV, no source was found to
have significant spectral
curvature, so each spectrum was described by a power-law model.
Each RoI is too small to allow characterizing by itself both the Galactic and
isotropic diffuse components, so the isotropic level was fixed to the
best fit over the entire sky and we left free the Galactic normalization only.

We used the Test Statistic $TS=2 \Delta \log \mathcal{L}$ (comparing the likelihood with and without the source) to quantify the
significance of sources. As for the 2FGL analysis, the detection threshold was set to $TS > 25$,
corresponding to a significance just over $4 \sigma$ for 4 degrees of freedom
(two for the localization, and two for the spectrum).
Sources below that threshold were discarded from the model, except for
the a priori known (at $>$100 MeV) extended
sources, which we retained to model the background even when they were not
clearly detected above 10 GeV.
No constraint was enforced on the minimum number of $\gamma$ rays for detected sources,
because above 10 GeV
and outside the Galactic plane the detection is not background limited.
In practice the faintest sources were detected with only 4 $\gamma$ rays.
We used binned likelihood functions as in 2FGL, handling $Front$ and $Back$ events
separately, with 0.05\degr and 0.1\degr spatial binning respectively,
and 10 energy bins per decade. 
At the end of the process 514 sources (including 18 extended sources) remained at $TS > 25$ among the 1705
input seeds.

\begin{figure}[t]
\begin{center}
\includegraphics[width=8cm]{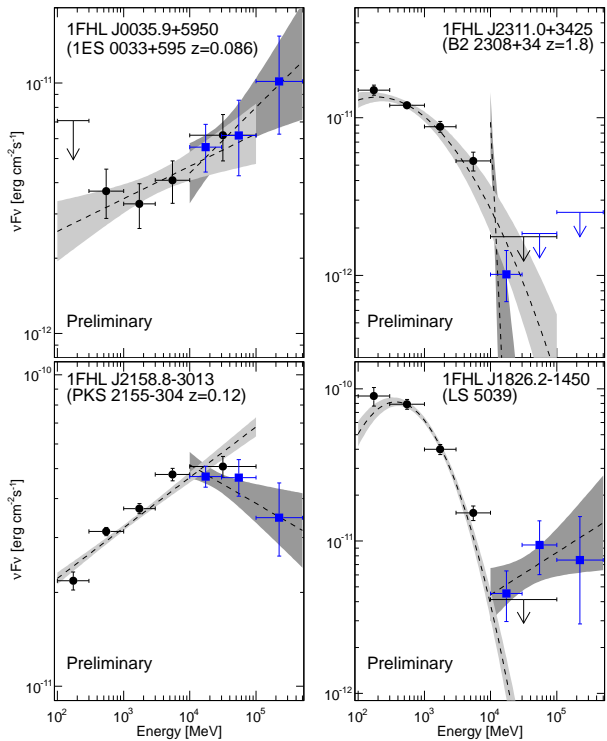}
\end{center}
\caption{
Spectra of four representative 1FHL sources with different spectral
shapes above 10 GeV:  the 1FHL sources associated with the blazars
1ES~0033+595 (z=0.086), PKS~2155-304
(z=0.117), B2~2308+34 (z=1.8), and with the high-mass binary system LS~5039.
The black points and light-grey bands depict the results
reported in the 2FGL catalog, while the blue data points and the
dark-grey bands depict the spectral results reported in this
work. See text for further details.
}
\label{ExampleSpectra1FHL} 
\end{figure}

Figure~\ref{ExampleSpectra1FHL} compares the spectral measurements reported in
the 2FGL paper (in the 100~MeV to 100~GeV energy range) with the results reported here
in the 10--500~GeV energy range, for four representative sources.  The blazar
1ES~0033+595 (z=0.086) has a 1FHL spectrum that is a continuation of the 2FGL
spectrum, while the classical TeV blazar PKS~2155$-$304 (z=0.117), which is a few times
brighter than the former, shows a clear turnover (from hard to
soft spectrum) at about 10~GeV. Given that PKS~2155$-$304 is a
relatively nearby source, this turnover must be due to an internal
break in the emission mechanism of this source. On the other hand, the
spectrum of the distant blazar B2~2308+34 (z=1.8) shows a clear cutoff (strong turnover) around 10 GeV which, given the high redshift of this
source, must be dominated by the absorption of $\gamma$-rays in the
extragalactic background light density (EBL). The fourth panel of
Figure~\ref{ExampleSpectra1FHL} shows the 1FHL spectrum of the
high-mass binary system LS~5039, which has a completely different shape with respect to
the 2FGL spectrum, indicating the presence of a new spectral component.
It is worth stressing that, when fitting the sources' spectral shapes
over this large dynamic range in energy spanning from 100 MeV to 100 GeV (as performed in the 2FGL), the highest
energies have a much lower statistical weight than the lower energies
because of the ($\sim$2 orders of magnitude) lower photon
count\footnote{
For a spectrum parametrized with a power-law function of energy with index $\alpha$, the
ratio between the number of photons above 100~MeV and the number of
photons above 10~GeV is given by $(10/0.1)^{\alpha - 1}$. For a
typical spectrum with $\alpha=2$, the number of detected photons above 10~GeV
would be a factor $\sim$30 less than the number above 100~MeV on
consideration of the relative acceptance of the LAT at these two
energies.}, and hence the best-fit spectral shape is determined
largely by the lower-energy $\gamma$ rays. 
Consequently, the spectral measurements at the highest energies ($>$10
GeV) can disagree with the 2FGL spectral fits for those sources that have 
turnovers, cutoffs, and/or new components arising at the highest
energies ($>$10~GeV), as illustrated in Figure \ref{ExampleSpectra1FHL}.
These deviations from the simple
spectral extrapolation from lower energies indicate the dominance of
other physical processes occurring at the source, 
or in the environment crossed by the $\gamma$ rays, and hence they are relevant for the proper understanding
of these sources. This is naturally one of the important motivations
for producing the 1FHL catalog.

\subsection{Associations}
\label{SrcAssociations}

The 1FHL sources were associated with (known) sources at other
wavelengths using similar procedures as for the 2FGL and the ``The
Second Catalog of Active Galactic Nuclei Detected by the Fermi Large
Area Telescope'' \cite[hereafter 2LAC,][]{LAT_2LAC}.

\begin{figure*} [th]
\begin{center}
\includegraphics[width=17.0cm]{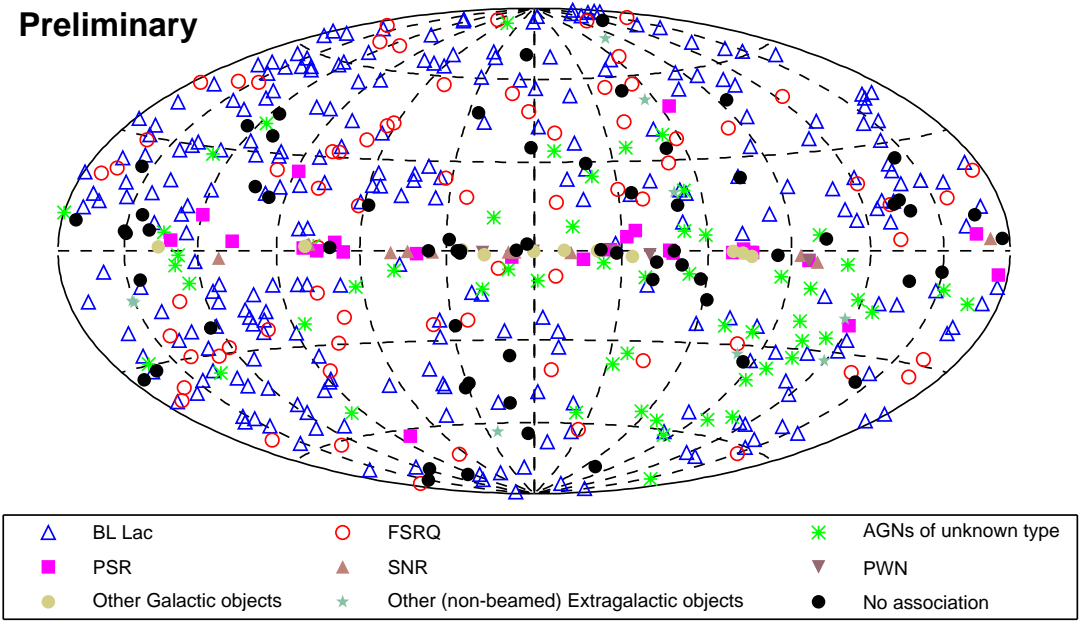}
\end{center}
\caption{\label{SkymapSummaryClasses} 
Sky map showing the sources by the source classes reported in Table \ref{TableSummarySrcClasses}}
\end{figure*}

Of the 514 sources in the 1FHL catalog, we found that 460 were already in
the 2FGL catalog, while 54 sources were not. 
Most of these 1FHL-but-non-2FGL sources  are located
outside the Galactic plane: 9 blazars, 8 blazar candidates, and a
large fraction of the 36 unassociated sources. 
The star-forming region called the Cygnus cocoon (in the Galactic plane) is also
part of this 54-source subset.

\begin{table}[t]
\begin{center}
\footnotesize
\caption{Summary of the 1FHL Source Classes}
\begin{tabular}{|l|c|c|c|}
\hline 
\textbf{Class Description} & \textbf{Number of} & \textbf{Fraction of}
& \textbf{Fraction of} \\
  & \textbf{Sources} & \textbf{Total\footnote{Fraction relative to the
      514 objects in the 1FHL catalog}} &
  \textbf{Associated\footnote{Fraction relative to the 449 objects
      with associated sources}}  \\
& & \textbf{[\%]} & \textbf{[\%]} \\
\hline 
Blazar BL Lac &  259 & 50.4 & 57.7 \\ 
Blazar FSRQ  &  71 &  13.8 & 15.8 \\
Unknown type AGN  & 55 & 10.7 & 12.2 \\
Pulsar   & 27 & 5.2 & 6.0 \\
Supernova remnant & 11 & 2.1 & 2.4 \\
Pulsar wind nebula & 6 & 1.2 & 1.3 \\
Other Galactic & 11 & 2.1 & 2.4 \\
Other extragalactic  & 9 & 1.8 & 2.0 \\
\hline
Unassociated source & 65 & 12.7 &  --- \\
\hline 
\end{tabular}
\label{TableSummarySrcClasses}
\end{center}
\end{table}

The sources that could be associated with sources of known nature (reported in
non-$\gamma$-ray catalogs) amount to 449. A brief summary of the statistics for the various source classes in
the 1FHL catalog is reported in Table~\ref{TableSummarySrcClasses}. 
A remarkable
characteristic of this catalog is that the blazars and blazar
candidates\footnote{The fraction of non-beamed AGNs is expected to be
only few percent, and hence most of the AGNs of unknown type are expected to
be blazars of either FSRQ or BL Lac type.} amount to $\sim$75\% of the entire catalog ($\sim$86\% of the
associated sources), indicating that this source class largely dominates
the highest-energy LAT sky.
The second largest source class is pulsars, with 5.2\% of the
catalog total. supernova remnants (SNRs) and pulsar wind nebulae (PWNe) together are only 4.5\% of
the catalog. Among the sources classified as {\it other extragalactic}, there
are radio galaxies, non-blazar active
galaxies, and the nearby galaxy Large Magellanic Cloud
 (LMC), which, given its proximity, has an extension of 2\degr; and the sources classified as {\it other Galactic} are binary
systems, globular clusters, and star forming regions, in addition to
6 sources that could be SNRs or PWNe.

From the  65 unassociated 1FHL sources, 5 are associated with extended (Galactic) unidentified H.E.S.S. sources, 25 are
associated with unidentified 2FGL sources (including 1 associated with
one of the 5 previously-mentioned Galactic H.E.S.S. unidentified sources), 5 are associated with unidentified 1FGL sources, and
2 are associated with unidentified sources from the 3$^{\rm rd}$ EGRET catalog \citep{3EGCatalog}.
The remaining 28 sources could not be associated with any $\gamma$-ray source reported previously. 
We note that the fraction of unassociated 1FHL sources is
only $\sim$13\% (65 out of 514), while  the fraction of 2FGL sources in the 1FHL catalog that 
could not be associated is only 6\% (25 out of 460). The
fraction of  unassociated sources reported in the 2FGL catalog was
$\sim$31\% (575 out of 1873).

The locations on the sky of the sources in the above-mentioned  classes are
depicted in Figure~\ref{SkymapSummaryClasses}.  To a good approximation,
the Galactic sources are located essentially in the Galactic plane
(apart from some pulsars), while
the blazars are distributed roughly uniformly outside
the Galactic plane.

\subsection{Characteristics of the 1FHL AGNs}
\label{FHL_AGNs}



Table \ref{TableSEDClassAndRedshift} summarizes the numbers of 1FHL AGN
associations belonging to the various SED classifications defined in
\citep{LATSEDs_BrightBlazars2010}, with and without
redshift determinations. Among all blazars, the dominant SED class is
{\it high-synchrotron-peaked} (HSP), which makes up for
$\sim$41\% of all the 1FHL AGNs. This is not a surprising result because
HSPs typically have a {\it hard} spectrum
(power-law index $\lapp$2) and hence they are expected to be the
AGN source class with the highest-energy photons. Table
\ref{TableSEDClassAndRedshift} 
also shows that the amount of AGNs with redshift in
the 1FHL catalog is 208 ($\sim$53\%), from which the fraction of
sources with redshift is 47\%, 46\% and 41\% for HSP, {\it intermediate-synchrotron-peaked} (ISP) and sources without
SED classification, and 76\% for the class {\it low-synchrotron-peaked} (LSP). The larger fraction of
LSPs with available redshifts is because 58 of the 99 LSPs  are actually FSRQs which, {\it by definition}, have their redshift measured.

\begin{table}[t]
\begin{center}
\footnotesize
\caption{Summary of SED Classifications and Available Redshifts for 1FHL Sources With AGN Associations}
\begin{tabular}{|c|c|c|}
\hline 
\textbf{SED } & \textbf{Number of } &
\textbf{Number of sources} \\ 
 \textbf{Classification} &   \textbf{Sources} & \textbf{with redshift} \\
\hline 
HSP  &  162 &  76 \\ 
ISP  &    61 &   28 \\
LSP  &    99 &   75 \\
Not Classified & 71 & 29 \\
\hline 
{\bf Total} & {\bf 393} & {\bf 208} \\
\hline 
\end{tabular}
\label{TableSEDClassAndRedshift}
\end{center}
\end{table}

\begin{figure*}[th]
\begin{center}
\includegraphics[width=8cm]{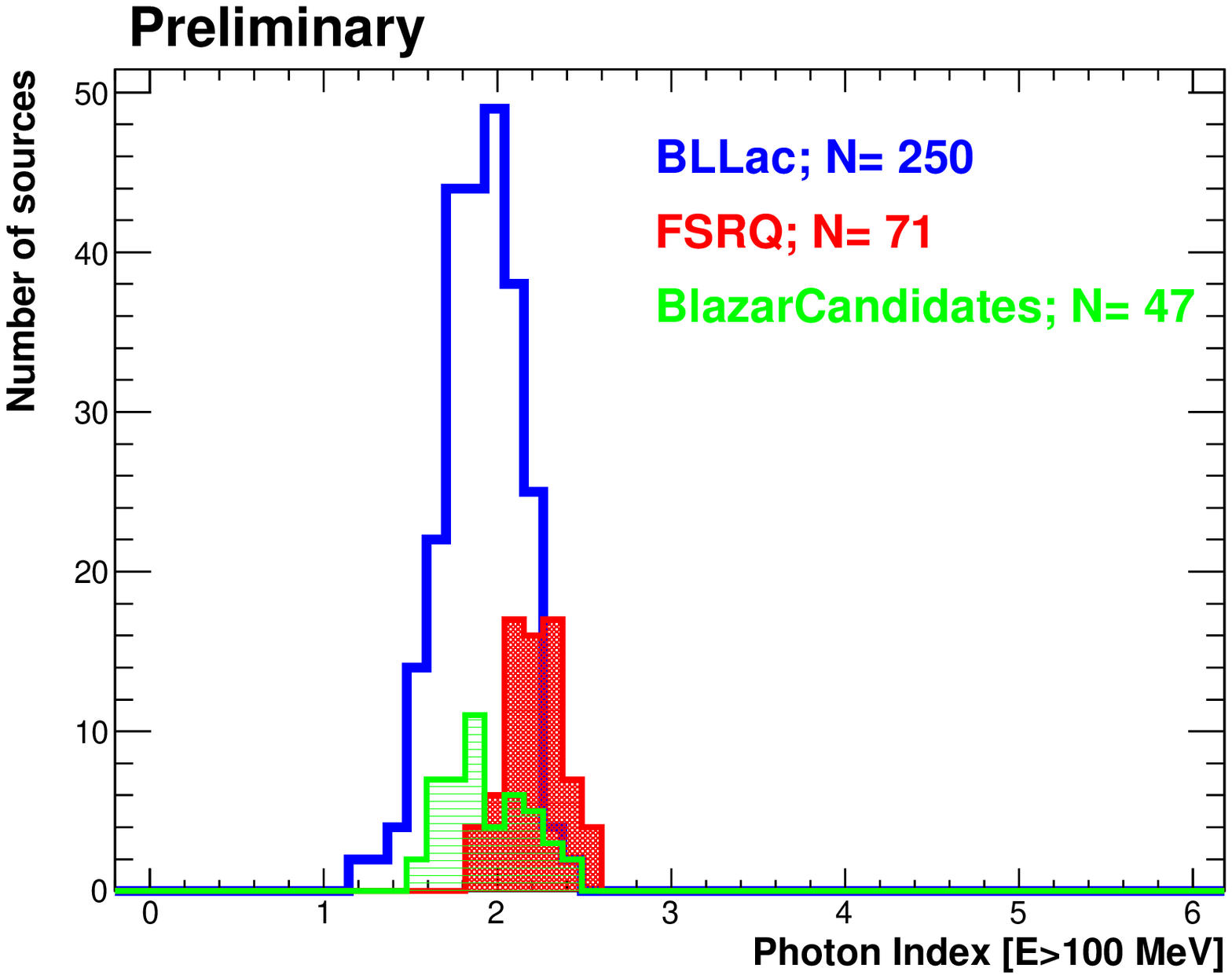}
\includegraphics[width=8cm]{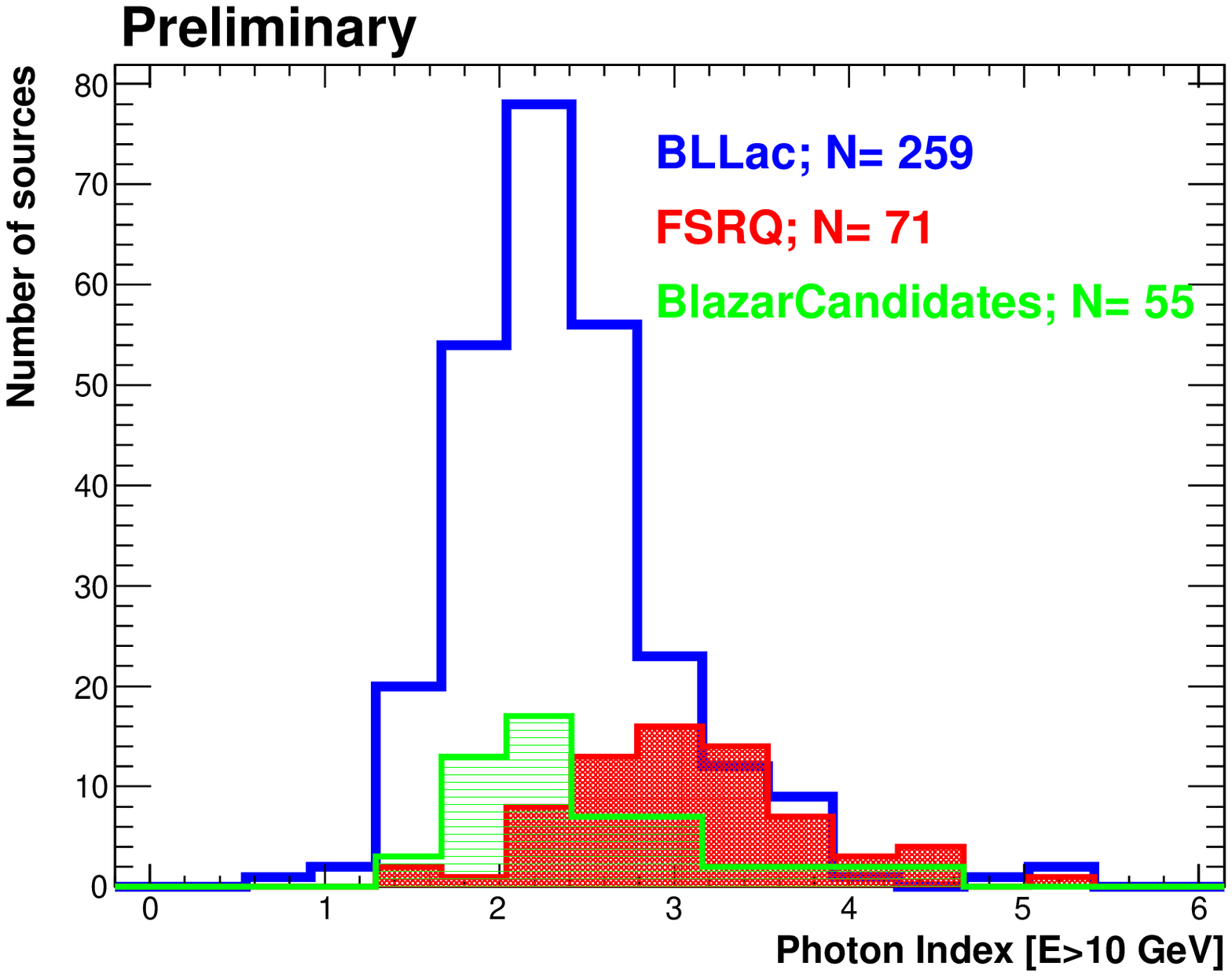}
\end{center}
\caption{\label{DistributionOfIndexAt100MeVAnd10GeV} 
Distribution of measured index for selected groups of 1FHL AGN sources above 100 MeV
({\it left}, extracted from the 2FGL catalog) and above 10 GeV 
({\it right}, \emph{this work}). The three
histograms show the distributions for three different groups of
AGN associations: BL Lacs (blue and not-filled histogram), FSRQs (red and
dotted-filled histogram), and blazar candidates (green and
horizontal-line filled histogram). See text for further details.}
\end{figure*}

Figure \ref{DistributionOfIndexAt100MeVAnd10GeV} shows the
distribution of the measured power-law indices of the 1FHL blazars in
the energy ranges 100 MeV to 100 GeV (extracted from the 2FGL
catalog table) and 10--500 GeV.
The figure does not show the nine 1FHL extragalactic sources that are associated with non-blazar AGNs. 
Note that the number of entries in the distributions in the left-hand panel
are less than in the right-hand panel. This is because
the 1FHL catalog
contains 17 AGN associations (9 BL Lacs and 8 blazar candidates) that do not exist in the 2FGL catalog
(\S~\ref{SrcAssociations}). 
The figure shows a clear softening in the spectra
of each source class when the minimum energy is increased from 100~MeV 
to 10~GeV. This is due to both intrinsic softening of the
spectra of many sources\footnote{The intrinsic softening can occur
  because of internal $\gamma$-$\gamma$ absorption, which is energy
  dependent, or because of a steep decrease with energy of the number of
  high-energy particles (presumedly electrons/positrons) that are
  responsible for the high-energy $\gamma$ rays.} 
  and the $\gamma$-ray
attenuation in the Extragalactic Background optical/UV Light (EBL) for distant ($z>0.5$)
sources. We also note that in both panels the photon indices of
the FSRQs cluster at the highest (softest) index
values, while BL Lacs have the lowest (hardest) values. 
So even when the spectra are characterized using photons above 10 GeV, we
find that about 30\% of the BL Lacs (77  out of 259) have indices harder than 2.
The index distribution of the blazar candidates (AGNs of unknown type)
is similar to that of BL Lacs, which suggests that a large
fraction of these blazar candidates are actually BL Lacs.

\begin{figure*}[th]
\begin{center}
\includegraphics[width=8cm]{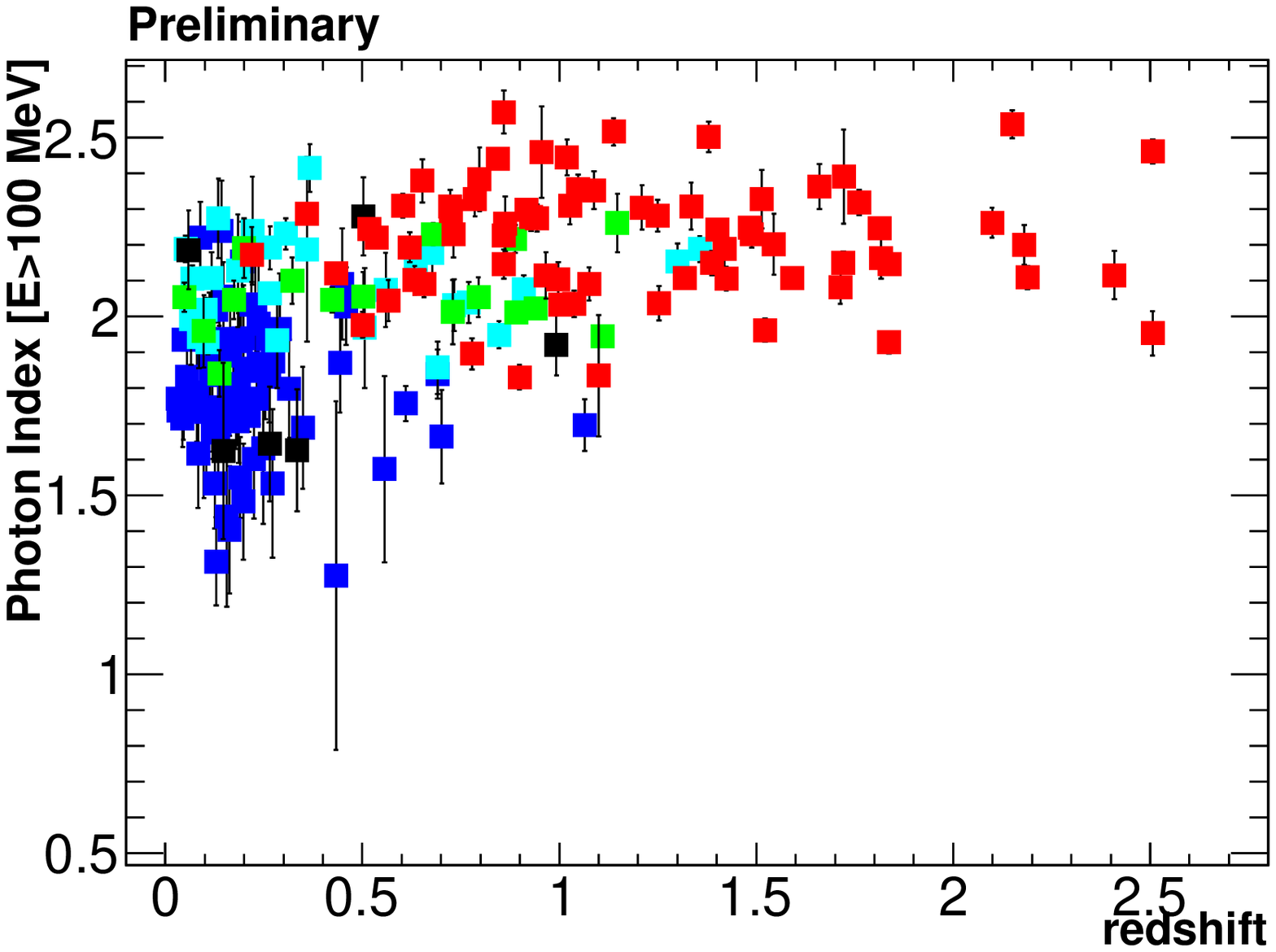}
\includegraphics[width=8cm]{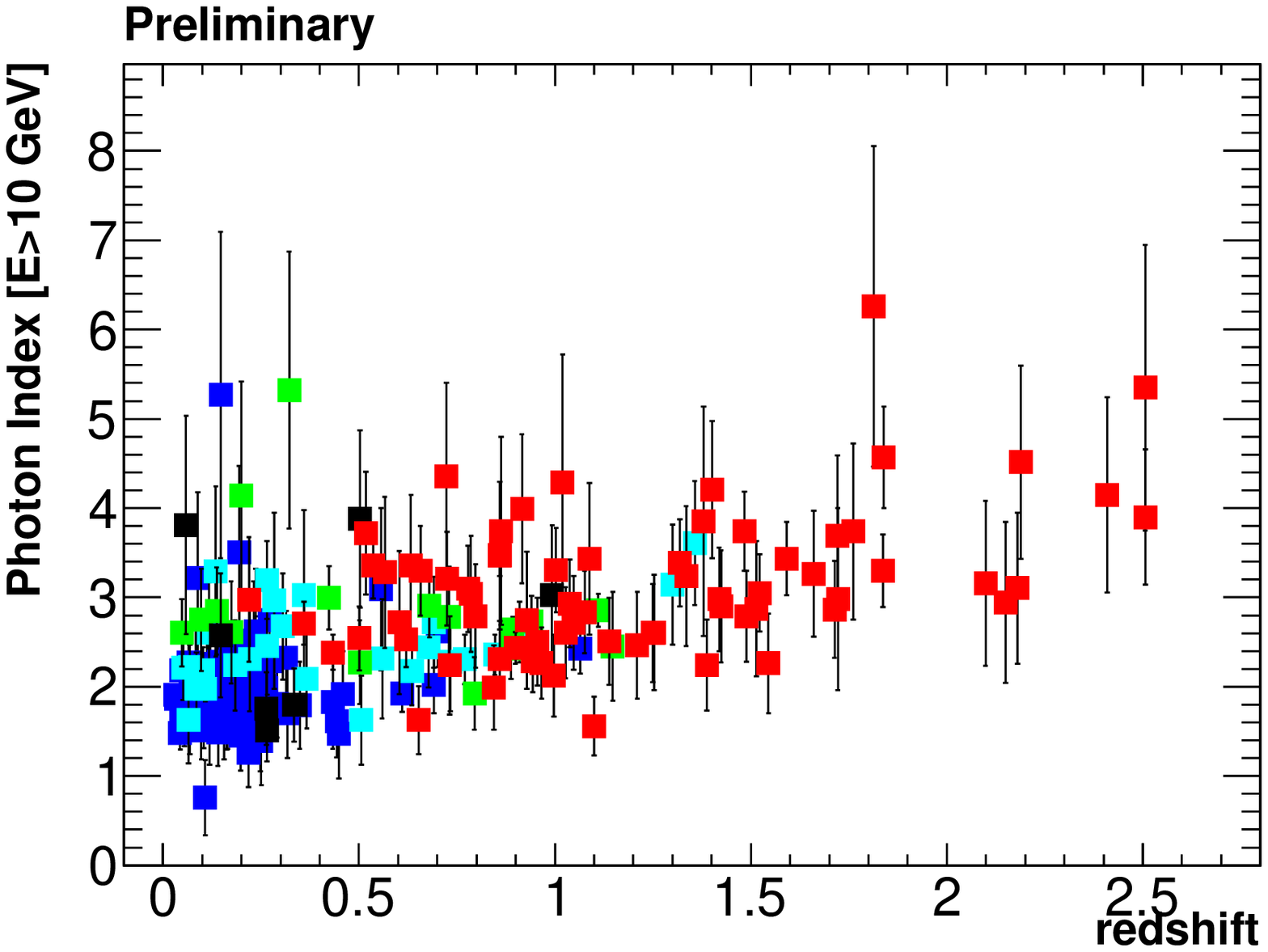}
\end{center}
\caption{\label{IndexVsRedshift} 
Power-law index versus redshift for the 1FHL sources with available
redshifts. The {\it left} panel shows the power-law index describing the
spectral shape above 100 MeV (extracted from the 2FGL catalog) and the
{\it right} panel shows the power-law index describing the spectral
shape above 10 GeV (\emph{this work}). In both panels, red indicates
FSRQs (71 sources), dark-blue for HSP-BL Lacs (73), 
light-blue for ISP-BL Lacs (27),  green for LSP-BL Lacs (16), and black for BL Lacs with unclassified SEDs (7).}
\end{figure*}

Figure \ref{IndexVsRedshift} shows a scatter plot of the photon index
($E>100$MeV and $E>10$ GeV) versus redshift for the various blazar subclasses: FSRQs, HSP-BL Lacs, 
ISP-BL Lacs, LSP-BL Lacs and BL Lacs without SED classification. 
There is no redshift evolution in the spectral shape characterized
with photon energies above 100 MeV, which is in agreement with the results reported in
Figure 19 of the 2LAC paper\footnote{The data used to produced the left
  panel from Figure~\ref{IndexVsRedshift} is actually the same used in
  the 2LAC, with the only difference being the selection of the blazar
sample: only 194 1FHL blazars (FSRQs+BL Lacs) are being used here.}. 
However, the photon index computed
with energies above 10 GeV depends on the redshift: the sources get softer with increasing
redshift. This trend is essentially invisible in the BL Lac sample,
which cluster at relatively low redshifts (mostly below 0.5);
but it is noticeable in the sample of FSRQs, which extends up to redshift 2.5.
A possible explanation for this redshift evolution in the spectral shape measured
with photon energies above 10 GeV (but not for the spectral shape computed with photon-energies
above 100 MeV) is the attenuation of the $\gamma$-rays in the optical/UV
density of photons from the EBL, which is energy dependent and only
affects photons above a few tens of GeV. 
In addition, one cannot exclude a potential cosmological evolution of
the FSRQ sample that introduces an intrinsic softening of the spectra. However, in order to be consistent with the
experimental observation reported in the left/right panels of Figure~\ref{IndexVsRedshift}, such
a  cosmological evolution of FSRQs should affect only
the emission above 10 GeV.

Because of the low photon count ($\sim$10 photons per source) above 10
GeV, quantifying the variability of the 1FHL sources is very challenging. 
We used the Bayesian Block algorithm from
\citep{Scargle1998}, which is unbinned in time, partitioning
the time series data into
piecewise constant segments (blocks), characterized by a rate (or flux) and duration.
Using the simulation results presented in \citep{Scargle2013}, an
acceptable fraction of false positives for detecting variability can
be easily specified. For the determination of the variability, we
considered a false positive threshold of 1\%.

Of the 514 1FHL sources, 43 are significantly variable. The variable sources are each associated with an AGN.
Among them, we find 22 LSPs
(22\% of the 99 1FHL LSP associations), 7 ISPs (13\% of the 61 1FHL ISP
associations), 6 HSPs (4\% of the 162 1FHL sources), and 8 sources with
no SED classifications (11\% of the 71 1FHL sources with unclassified SEDs).
One of the noteworthy characteristics is that 
most of the 1FHL sources identified as variable belong to the blazar subclass
LSP, 
not to the dominant HSP subclass, which also has a larger number of high-energy photons. 
It is worth noting that the three classic VHE blazars with the highest
measured variability above a few hundred GeV, namely Mrk~421, Mrk~501 and
PKS~2155$-$303 (which are HSPs) are not found to be variable in the 1FHL catalog. This is
surprising, given that these three also have the largest numbers of detected photons
above 10 GeV:  432, 247, and 132, respectively.
Moreover, the fraction of 1FHL LSPs identified as variable
($\sim$22\%) is substantially
higher than the fraction of 1FHL HSPs identified as variable ($\sim$4\%). 
This trend was already observed in the 2FGL blazars at energies
above 100 MeV, and reported in the 2LAC paper (e.g., see Figs. 26 and 27 of that work). 
Therefore, we can confirm that across the entire energy range of the
LAT, the LSPs are more variable than the
HSPs.  These experimental observations show that the variability in the falling segment of the high-energy (inverse
Compton) SED bump is greater than that in the rising segment of the SED bump.

\subsection{Good Candidates for VHE Detection}
\label{GoodVHE}

As mentioned above, the number of known VHE sources is $\sim$100, 
which is a low number in comparison with the $\sim$1000 \gray\ sources
reported by \FermiLAT, and the many thousands reported at
radio/optical/X-ray frequencies. 
This low number
of VHE sources precludes detailed population studies, which are crucial
to understand the physical properties of the various source types and
move toward unification schemes.

The most sensitive instruments to perform VHE astronomy are the
Imaging Atmospheric Cherenkov Telescopes (IACTs), 
which detect \grays\ through the observation of the (extended air)
electromagnetic shower 
induced by the \gray\ in the Earth atmosphere. 
The main reasons for this very low number of known VHE sources (mostly
discovered by IACTs) 
are (a) the narrow field of view ($3^{\circ}-5^{\circ}$)
of IACT cameras, hence requiring many pointings to scan a region of
the sky, (b) the $\sim$10\% operational duty cycle ($\sim$1000 hours annually), because IACTs
  can only operate on moonless nights with good weather conditions,
  and   (c) the low photon count at VHE, hence requiring long
observing times.
In order to increase the efficiency of the searches for new
  sources at VHE with IACTs, one can exploit the all-sky
  coverage capability, the large duty cycle,
  and the high \gray\ energy band of \FermiLAT to guide IACT observations toward these sources
  with high chances of being VHE emitters.

We identified 213 1FHL sources as  
 good candidates for being VHE emitters.
and be potentially detected with the current generation of ground-based \gray\
instruments. 
This list consists of 1FHL sources that have not been detected at VHE
but have properties similar to the 84 1FHL sources that have associations with known
VHE sources, and is based on the following selection criteria: 
(a)  Index10GeV $<$ 3; (b) Sig30GeV $>$3;  and (c) F50GeV $> 10^{-11}$ ph cm$^{-2}$ s$^{-1}$.
The parameter Index10GeV is the spectral index above 10~GeV, which is directly
one of the parameters from the spectral fits, while F50GeV is the
estimated flux above 50~GeV, which is a derived quantity from the
spectral fits above 10~GeV.
The parameter Sig30GeV is the {\it pseudo significance} of the signal
above 30~GeV, which is a derived quantity computed as
$\sqrt{TS30\_100GeV+TS100\_500GeV}$, 
where TS30\_100GeV and TS100\_500GeV are the test statistic values 
for the energy bands 30--100~GeV and 100--500~GeV, respectively. 
Although the three parameters used above are highly correlated, the best indicator of the ``VHE-detectability'' is F50GeV;
the greater the estimated flux above 50~GeV with \FermiLATc, the higher
the VHE flux, and hence the easier to be detected with ground-based \gray\ instruments.

A subset of the above-mentioned list of VHE source candidates, namely the ones with  F50GeV $>
3 \cdot 10^{-11}$ ph cm$^{-2}$ s$^{-1}$, was released to all major IACTs in
September 2012.
One of the interesting characteristics about this list of VHE
candidate sources 
is the presence of 18 high-redshift ($z$\gapp 0.2) blazars.
High-redshift sources are important  for two reasons, \emph{(i)}
population studies related to the potential cosmological evolution of the
\gray\ emission of blazars, and \emph{(ii)} studies related to the
properties of objects, and radiation fields that are
located between us and the blazars at cosmological distances. 
Many of these sources will be observed and detected by all
IACTs over the coming years.
Every new VHE
discovery implies that the known VHE sky increases by  $\sim$1\%, and
the high-redshift sources, if/when detected with IACTs, will expand 
the volume of the known Universe at VHE, hence enabling the
  study of their cosmological evolution.

\bigskip 
\begin{acknowledgments}

The $Fermi$ LAT Collaboration acknowledges support from a number of
agencies and institutes for both development and the operation of the
LAT as well as scientific data analysis. These include NASA and DOE in
the United States, CEA/Irfu and IN2P3/CNRS in France, ASI and INFN in
Italy, MEXT, KEK, and JAXA in Japan, and the K.~A.~Wallenberg
Foundation, the Swedish Research Council and the National Space Board
in Sweden. Additional support from INAF in Italy and CNES in France
for science analysis during the operations phase is also gratefully
acknowledged.

\end{acknowledgments}

\bigskip 

\end{document}